\documentclass[twocolumn,showpacs,amsmath,amstex,amssymb,mathfonts,prb]{revtex4-1}
\usepackage{graphicx,bm,units}
\usepackage{times}
\usepackage{color}

\usepackage{color}

\begin{document}

\title{Band mass anisotropy and the intrinsic metric of fractional quantum Hall systems}
\author{Bo Yang$^1$, Z. Papi\'c$^2$, E. H. Rezayi$^3$, R. N. Bhatt$^{2}$, and F. D. M. Haldane$^1$}
\affiliation{$^1$ Department of Physics, Princeton University, Princeton, NJ 08544, USA}
\affiliation{$^2$ Department of Electrical Engineering, Princeton University, Princeton, NJ 08544, USA}
\affiliation{$^3$ Department of Physics, California State University, Los Angeles, California 90032, USA}

\date{\today}

\begin{abstract} 
It was recently pointed out that topological liquid phases arising in the fractional quantum Hall effect (FQHE) are not required to be rotationally invariant, as most variational wavefunctions proposed to date have been. Instead, they possess a geometric degree of freedom corresponding to a shear deformation that acts like an intrinsic metric. We apply this idea to a system with an anisotropic band mass, as is intrinsically the case in many-valley semiconductors such as AlAs and Si, or in isotropic systems like GaAs in the presence of a tilted magnetic field, which breaks the rotational invariance. We perform exact diagonalization calculations with periodic boundary conditions (torus geometry) for various filling fractions in the lowest, first and second Landau levels.  In the lowest Landau level, we demonstrate that FQHE states generally survive the breakdown of rotational invariance by moderate values of the band mass anisotropy. At 1/3 filling, we generate a variational family of Laughlin wavefunctions parametrized by the metric degree of freedom. We show that the intrinsic metric of the Laughlin state adjusts as the band mass anisotropy or the dielectric tensor are varied, while the phase remains robust. In the $n=1$ Landau level, mass anisotropy drives transitions between incompressible liquids and compressible states with charge density wave ordering. In $n\geq 2$ Landau levels, mass anisotropy selects and enhances stripe ordering with compatible wave vectors at partial 1/3 and 1/2 fillings.
\end{abstract}


\maketitle \vskip2pc

\section{Introduction}\label{sec_introduction}

Two-dimensional electron systems (2DES) placed in a high magnetic field exhibit a wide variety of strongly correlated phases, which
have been the subject of numerous theoretical and experimental investigations since the first observation of 
fractionally quantized Hall conductivity~\cite{tsg}. Examples of such phases of matter are the Laughlin states~\cite{laughlin}, describing partial fillings $\nu=1/3,1/5$ of the lowest ($n=0$) Landau level (LL), as well as their generalizations to other odd-denominator fillings in the framework of hierarchy~\cite{prange} and composite fermion theory~\cite{jainbook}. These phases are topologically ordered and possess quasiparticle excitations with fractional statistics. At half filling of the first excited, $n=1$ LL, an even more exotic paired state, the Moore-Read Pfaffian~\cite{mr}, might be realized, which possesses non-Abelian excitations -- the Majorana fermions~\cite{mr, readgreen}. 

Besides the incompressible liquids, some fillings $\nu$ also lead to compressible phases without 
quantized conductance. This is the case with the simplest of all fractions -- $\nu=1/2$ in $n=0$ LL -- which 
is a Fermi liquid of composite fermions~\cite{hlr}, that only supports an anomalous Hall effect~\cite{anomalous}.
Generically for any $\nu$, apart from the incompressible liquids, the natural candidates are compressible phases 
that break translational symmetry, such as charge density waves (CDWs)~\cite{anisotropic}. Those were in fact proposed to
describe the ground state of 2DES before FQHE was observed~\cite{fukuyama}. When $\nu$ is very small (under $1/7$),
a correlated Wigner crystal indeed becomes energetically superior to a Laughlin-type state~\cite{lamgirvin, edduncanky_wigner}. 
Furthermore, when $\nu > 2$, several varieties of states with broken translational symmetry become energetically 
favorable. Around half filling of $n\geq 2$ LL, the ground state becomes a charge density wave in one spatial direction or a ``stripe"~\cite{kfs_prl, kfs_prb, moessner_chalker, edduncanky_stripe};
away from half filling, two-dimesional crystalline order sets in, resulting in a ``bubble" phase~\cite{kfs_prl,kfs_prb,moessner_chalker, edduncanky_bubble}.    
Some of these phases also occur in $n<2$ LLs when the hard-core component of the effective interaction
is significantly softer than Coulomb~\cite{rh_pf}. More recently, the experiments have shown~\cite{xia2011} that it is possible to have, at the same time, the quantization of resistance and anisotropic transport, suggesting a possible coexistence of an incompressible liquid with a compressible (``nematic") phase~\cite{mulligan}.  

Theoretical understanding of the FQHE was pioneered by Laughlin's method of many-body trial wavefunctions~\cite{laughlin}. Model wavefunctions can be formulated using the conformal field theory~\cite{mr}, and
conveniently evaluated in finite-size systems via exact diagonalization of the parent Hamiltonians~\cite{haldane_prange}. 
In addition, excitation spectra containing quasiparticles/quasiholes above the ground state can be studied. 
Such analytical and numerical studies are made much easier by exploiting symmetry
and the corresponding quantum numbers to characterize the ground state and excitations. 
To this end, rotational symmetry has been very useful~\cite{haldane_sphere}; ocassionally, periodic boundary conditions have also been used~\cite{yhl, duncan_translations}.

However, as it was recently pointed out~\cite{duncan}, rotational symmetry is not fundamental to the appearance of FQHE. In a theoretical treatment of the FQHE, it is important to distinguish several ``metrics" that naturally arise in the problem. The band mass tensor yields a metric that defines the shape of the LL orbitals. A second metric derives from the dielectric tensor of the semiconducting material, and defines the shape of the equipotential contours around an electron. 
Rotational invariance means that these two metrics are congruent, however in a real sample they might be different from one another, thus lifting the rotational invariance. 

It turns out, however, that a given FQH state also possesses an intrinsic metric that is derived from the two types introduced above. FQH fluids can be described as condensates of composite bosons~\cite{zhk}, which are topological objects that explain the quantization of the Hall conductance and the emergence of fractionally-charged quasiparticles. However, apart from topology, composite bosons also have a geometrical degree of freedom -- the intrinsic metric that controls their \emph{shape}. In a rotationally invariant case, the intrinsic metric is equal to the metric in the Hamiltonian; more generally, as we explicitly demonstrate below, the shape of composite bosons can be defined even in systems without rotational invariance. The fluctuation of the intrinsic metric plays an important role for the geometrical field theory of the FQHE~\cite{duncan_unpublished}, and determines the	 energetics of quasiparticles, collective modes etc. Generalizations of the Laughlin wavefunction to the broken-rotational-symmetry case have been proposed for liquid crystal and nematic Hall phases~\cite{ciftja}, and very recently Laughlin and Moore-Read wavefunctions (as well as their parent Hamiltonians) have been formulated for the anisotropic case~\cite{kunyang_aniso}.

The motivation for studying the effect of anisotropy in FQHE is twofold. On the one hand, anisotropy probes the variations of the intrinsic metric of FQH fluids, a fundamental physical quantity that relates to the geometric description of FQHE. Secondly, we explore the possible effects resulting from tuning the rotational-symmetry breaking by an external parameter. Note that the rotational invariance is explicitly broken in real samples due to the presence of impurities, which are essential for the emergence of FQH plateaus. Furthermore, it is possible to induce the breaking of rotational invariance by tilting the magnetic field~\cite{tilt}, or by using systems with anisotropic bands, e.g. many-valley semiconductors like AlAs or Si in the presence of uniaxial stress. The former method is performed routinely and belongs to the most popular techniques for studying the FQHE; the latter method is  relevant to AlAs~\cite{shayegan} and some new classes of materials where FQHE might be studied. We provide brief arguments how the tilting of the field can be mapped to an effective variation of the metric, and then focus on the second method.

This paper is organized as follows. In Sec.~\ref{sec_model} we introduce and motivate the model for a FQH system with band mass anisotropy. In Sec.~\ref{sec_laughlin} we discuss the intrinsic metric of the Laughlin state. We define a family of the Laughlin wavefunctions characterized by the varying shape of their elementary droplets. Intrinsic metric is determined variationally by optimizing the overlap between this family of wavefunctions and the exact Coulomb ground state. In Sec.\ref{sec_ll1} we perform exact diagonalization of finite systems at several filling factors to explore quantum phase transitions that occur as a function of the anisotropy. Our conclusions are presented in Sec.\ref{sec_conclusion}. 

\section{Model}\label{sec_model}  

Consider an electron moving in the plane with the perpendicular magnetic field $B\hat{z}=\nabla\times \mathbf{A}(\mathbf{r})$. The Hamiltonian can be written in the following, manifestly covariant form:
\begin{eqnarray}
K^{\alpha,\phi} =\frac{1}{2m} g^{ab} \pi_{a} \pi_{b}.
\end{eqnarray}
Here $\pi_a= p_a - \frac{e}{c} A_a(\mathbf{r}), a=x,y$ represents the dynamical momentum, and $g$ is the mass tensor parametrized by the anisotropy $\alpha$ and the angle of the principal axis $\phi$. The mass tensor is unimodular $\det g=1$. In the isotropic case when $g$ is the unit matrix, we can obtain the single-particle energies (Landau levels) by choosing, for example, a symmetric gauge $A_x=By/2, A_y=-Bx/2$. In this case, the dynamical momenta become 
$\pi_x=-i\hbar \frac{\partial}{\partial x}+\frac{\hbar}{2\ell_B^2}y$ and $\pi_y=-i\hbar \frac{\partial}{\partial y} - \frac{\hbar}{2\ell_B^2}x$, in terms of the magnetic length $\ell_B=\sqrt{\hbar/eB}$. The Hamiltonian can be transformed into diagonal form $K=\frac{\hbar\omega_c}{2} \left(a^\dagger a + \frac{1}{2} \right)$ with the help of ladder operators $a \propto \pi_x + i\pi_y$ and $a^\dagger \propto \pi_x - i \pi_y$.
However, for each value of $a^\dagger a$, there is residual degeneracy equal to the number of the magnetic flux quanta $N_\phi$. This degeneracy is resolved by a second pair of operators $b, b^\dagger$ that commute with $a, a^\dagger$ and depend on the \emph{guiding center} coordinates of the electron, $R^a = r^a - \frac{\epsilon^{ab}}{\hbar} \pi_b \ell_B^2$. Operators $b^\dagger$ create the (unnormalized) single particle eigenstates of the lowest LL, 
\begin{equation}
\phi_l^{\alpha=1, \phi=0} (z) = z^l e^{-z^* z/4\ell_B^2},
\end{equation}\label{singleparticleorbital}
with $z=x+iy$ being the complex coordinate of an electron in the plane (and $z^*$ denoting its complex-conjugate). The quantum number $l$ is an eigenvalue of the angular momentum $L_z$ and the single particle states $\phi_l$ are localized on concentric rings around the origin. 

To illustrate the effect of mass anisotropy, we take the principal axes of the mass tensor to be along the $x$ and $y$ directions ($\phi=0$), with different masses along the two directions ($\alpha \neq 1$).  Via simple rescaling $x\rightarrow x/\sqrt{\alpha}, y\rightarrow y\sqrt{\alpha}$, and therefore introducing
$\tilde{a} \propto \sqrt{\alpha} \pi_x + \frac{i}{\sqrt{\alpha}}\pi_y$ and $\tilde{a}^\dagger \propto  \sqrt{\alpha} \pi_x - \frac{i}{\sqrt{\alpha}}\pi_y$,
we can immediatealy write down the single particle orbitals for this case:
\begin{equation}\label{singleparticleorbitalg}
\phi_l^{\alpha} (x,y) = \left( \frac{x}{\sqrt{\alpha}} + iy\sqrt{\alpha} \right)^l e^{-\frac{1}{4\ell_B^2} \left( \frac{x^2}{\alpha} + \alpha y^2 \right)}.
\end{equation}
Notice that the probability density $|\phi_l|^2$ is no longer localized on a circle, but rather an ellipse for $\alpha \neq 1$. Therefore, on a single particle level, the effect of mass anisotropy is to stretch or squeeze the one-body orbitals along certain directions, possibly rotating the principal axis (for $\phi \neq 0$).

As we mentioned in Sec.~\ref{sec_introduction}, certain semiconductor materials are likely to have non-trivial metric defined by the anisotropy. Alternatively, the effective mass tensor can be experimentally tuned by tilting the magnetic field~\cite{xia2011}. Tilting is known to produce complicated effects because it induces the coupling between electronic subbands and LL mixing, and a detailed analysis will be presented elsewhere. However, with a normal confinement (perpendicular to the Hall surface) given by a harmonic well, the Hamiltonian with a strong perpendicular magnetic field $B$ and a non-zero in-plane magnetic field $B_\parallel$ can be solved exactly, yielding two characteristic harmonic oscillator frequencies~\cite{demler}:
\begin{eqnarray}
\omega_{1,2}^2=\frac{1}{2}[(\omega_z^2+\omega_c^2)\pm(\omega_z^2-\omega_c^2)\sec(2\bar{\theta})],
\end{eqnarray}
where $\omega_z$ is the harmonic frequency of the normal confinement. The mixing of the cyclotron frequency and the confinement frequency is parametrized by $\tan 2\bar{\theta}=2\omega_c\omega_\parallel/(\omega_c^2-\omega_z^2)$, where $\omega_\parallel=eB_\parallel/m$. Writing $\lambda_{i=1,2}=\omega_i/\omega_c$, the anisotropy parameter becomes $\alpha=\lambda_1\lambda_2/(\lambda_1\sin^2\bar{\theta}+\lambda_2\cos^2\bar{\theta})$. Parametrizing the in-plane magnetic field by $\mathbf{B}_\parallel=(B_\parallel\cos\phi,B_\parallel\sin\phi)$, the effective metric associated with the tilt is given by
\begin{eqnarray}
\nonumber g=\left(\begin{array}{ccc}
\cosh2\theta+\sinh2\theta\cos2\phi &\sinh2\theta\sin2\phi\\
\sinh2\theta\sin2\phi & \cosh2\theta-\sinh2\theta\cos2\phi\end{array}\right), \\ 
\end{eqnarray}
where $\cosh 2\theta=\frac{1}{2}(\alpha+\frac{1}{\alpha})$. Therefore, the effect of tilting on the LLL single-particle levels can be captured by the  variation of the mass tensor. 

In order to study a finite, interacting system of $N_e$ electrons, it is convenient to choose a compact surface to represent the 2DES. As we emphasized in Sec.\ref{sec_introduction},  the presence of mass anisotropy destroys rotational invariance, and one must use periodic boundary conditions~\cite{yhl, duncan_translations} i.e. put the 2DES on the surface of a torus. The unit cell can generally be chosen as a parallelogram with sides $\mathbf{a}$ and $\mathbf{b}$ whose area $\mathcal{S}$ is quantized because of the magnetic translations algebra: $\mathcal{S}\equiv |\mathbf{a} \times \mathbf{b}| = 2\pi \ell_B^2 N_\phi$. The single-particle states compatible with periodic boundary conditions are given in the Landau gauge by
\begin{eqnarray}\label{landaugauge}
\nonumber \phi_{j,n}^\alpha (\mathbf{r}) = \frac{1}{\sqrt{\mathcal{N}}} \sum_k e^{i\left(X_j+ka\right)y -\frac{1}{2\alpha}\left(X_j + ka + x \right)^2} \\
H_n \left( \frac{X_j+ka+x}{\sqrt{\alpha}} \right),
\end{eqnarray}
where $j=0,\ldots,N_\phi-1$, $X_j \equiv 2\pi j/b$, normalization factor is $\mathcal{N}=b\sqrt{\pi} \sqrt{\alpha}2^n n!$, and the sum over $k$ extends over all integers. We have set $\ell_B=1$. The wavefunction for the $n$th LL involves a Hermite polynomial $H_n$. For simplicity, we assumed the case of a rectangular torus and $g={\rm diag}\left[\alpha, 1/\alpha\right]$, but Eq.(\ref{landaugauge}) can be generalized to an arbitrary shape/anisotropy using Jacobi theta functions.

Many-body states, like in the isotropic case, can be classified using a crystal quasimomentum $\mathbf{K}$~\cite{duncan_translations} defined in a Brillouin zone. With a suitable definition of the Brillouin zone, incompressible states always occur at $\mathbf{K}=0$, and are characterized by the gap in their excitation spectrum.  The Hamiltonian for $N_e$ electrons is given by the sum of the kinetic term and the Coulomb interaction,
\begin{eqnarray}\label{fullham}
H^{\alpha,\phi} = \sum_i K_i^{\alpha,\phi} + \sum_{i<j} \frac{1}{|\mathbf{r}_i - \mathbf{r}_j|_\epsilon}.
\end{eqnarray}
2DES is embedded in a three-dimensional dielectric host material, which is characterized by its own dielectic tensor $\epsilon$, 
defining the metric for the distance $|\mathbf{r}_i-\mathbf{r}_j|_\epsilon$ between interacting electrons. This tensor in general can be
different from the one that parametrizes the cyclotron orbits ($\alpha,\phi$). However, physical properties are determined only by the relative difference
 between the mass tensor and the dielectric tensor, and for simplicity we can set the latter to unity. In other words, we assume that Coulomb interaction is isotropic in space, hence its Fourier transform
is $V(\mathbf{q}) = 1/q \equiv 1/\sqrt{q_x^2+q_y^2}$ (to model finite-width effects, we use the softened form of $V(\mathbf{q})$, following
the Fang-Howard prescription).
Projected to a single $n$th LL, the interaction part of the Hamiltonian becomes
$H=\sum_{\{j_i\}}V_{j_1j_2j_3j_4}c_{j_1}^\dagger c_{j_2}^\dagger c_{j_3}c_{j_4}$,
where 
\begin{eqnarray}\label{matel}
\nonumber V_{j_1j_2j_3j_4} = \frac{1}{2\mathcal{S}} \sum_{\mathbf{q}}^{'} V(\mathbf{q}) \mathcal{L}_n \left( \frac{1}{2} q_g^2 \right)  \\
 e^{-\frac{1}{2} q_g^2 } e^{iq_x \left( X_{j_1}-X_{j_3}\right)} \delta_{t,j_1-j_4}^{'} \delta_{j_1+j_2,j_3+j_4}^{'},
\end{eqnarray}
where $q_g^2 \equiv g^{ab}q_a q_b$ (e.g. in case of a diagonal mass tensor, $q_g^2=\alpha q_x^2 + q_y^2/\alpha$) and $\mathcal{L}_n$ is the Laguerre polynomial. The primed $\delta$-functions are to be taken $({\rm mod} \; N_\phi)$, and the sum over $\mathbf{q}$ extends over the reciprocal space (the prime on the sum indicates that the diverging $\mathbf{q}=0$ term is cancelled by the positive background charge). 

Apart from the many-body translational symmetry, discrete symmetries can be used to further reduce the Hilbert space. Several types of Bravais lattices are possible, depending on the angle $\theta$ between the sides of the torus, $\mathbf{a}$ and $\mathbf{b}$, and the aspect ratio, $|\mathbf{a}|/|\mathbf{b}|$. Both of these can be tuned as free parameters. In the presence of anisotropy, however, the highest symmetry is only given by the rectangular lattice, even when $|\mathbf{a}|=|\mathbf{b}|$. Tuning the angle $\theta$ enables to perform the area-preserving deformations of the torus, which is useful in resolving the collective modes of FQH states in finite systems, and probing quantities such as Hall viscosity~\cite{duncan}.

In this paper we only consider spin polarized electrons and neglect the so-called multicomponent degrees of freedom, which can be the usual spin or bilayer/valley degree of freedom. This means that the filling factors we refer to as $n+\nu$ correspond to $kn+\nu$ in experiments, where integer $k$ denotes the additional degeneracy that comes from several ``flavors" of electrons.

\section{Anisotropy in the lowest Landau level: robustness and the intrinsic metric of the Laughlin state} \label{sec_laughlin}

\begin{figure}[ttt]
 \includegraphics[width=0.7\linewidth,angle=270]{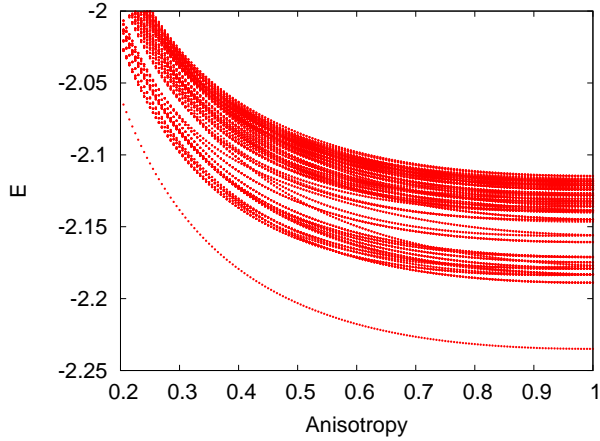}
\caption{(Color online). Energy spectrum in units of $e^2/\epsilon l_B$ as a function of anisotropy $\alpha$ for the square unit cell and $n=0$ LL Coulomb interaction at $\nu=1/3$. The system is $N_e=7$ electrons and $\phi=0$. Due to the square unit cell, the spectrum is symmetric under $\alpha \rightarrow 1/\alpha$.}
\label{fig_onethirdenergy}
\end{figure}

In Fig.~\ref{fig_onethirdenergy}, we present the energy spectrum of the Coulomb interaction at $\nu=1/3$ as a function of anisotropy (we assume $\phi=0$). The system is placed on the torus with a square unit cell, and energies are expressed in units of $e^2/\epsilon\ell_B$. A very flat minimum around isotropy point and the existence of a robust gap suggest that the ground state of the Coulomb interaction at $\nu=1/3$ is remarkably stable to variation in anisotropy. As we show below, in this range of $\alpha$, the ground state is described by a generalized Laughlin wavefunction. Moreover, the set of lowest neutral excitations, forming a magneto-roton branch, are also stable and separated from the rest of the spectrum. Within this manifold, some level crossings occur as $\alpha$ is changed, but this only corresponds to the redistribution of the levels within a roton branch. Beyond $\alpha \approx 0.5$, the ground-state energy rises, indicating an instability and the eventual destruction of the Laughlin phase.   

In rotationally-invariant situations, the incompressible liquids at fillings $\nu=1/m$ of $n=0$ LL ($m$ being an odd integer) are described by the Laughlin wavefunction~\cite{laughlin}. In the geometry of an infinite plane, the Laughlin state is given by 
\begin{equation}\label{laughlinwf}
\Psi_L^{\nu=1/m} = \prod_{i<j}(z_i-z_j)^m e^{-\sum_k z_k^* z_k/4\ell_B^2}.
\end{equation}
Here $z=x+iy$ stands for the usual complex representation of the coordinates in the plane, and can also be expressed in terms of spinor coordinates providing a mapping to the spherical geometry~\cite{haldane_sphere}. However, the Laughlin state can also be extended to the torus geometry~\cite{laughlin_torus}, where continuous rotation symmetry is broken down and survives at most in form of a discrete subgroup. In this case, $\Psi_L^{\nu=1/m}$ is defined by its short-distance correlations which assume the form of the (odd) Jacobi  $\vartheta_1$ theta function of $z_i - z_j$~\cite{laughlin_torus}. 

As a trial wavefunction, $\Psi_L^{\nu=1/3}$ provides an excellent description of the physical system at $\nu=1/3$ in the limit of strong cyclotron energy with respect to the Coulomb repulsion in the 2DES plane, $\hbar \omega_c \gg e^2/\epsilon \ell_B$, when
excitations to higher LLs are prohibited. In this limit, the operators $a,\tilde{a}$ act trivially and the only dynamical degrees of freedom
are the (non-commuting) guiding centers. Up to the normalization, we can then view the wavefunction (\ref{laughlinwf}) as follows
\begin{equation}\label{laughlinwfg}
\Psi_L^{\nu=1/m} (g) \propto \prod_{i<j} \left( b_i^\dagger (g) - b_j^\dagger (g) \right)^m |0\rangle,
\end{equation}
where $b_i^\dagger (g)$ explicitly depend on the metric $g$~\cite{duncan}. For general $g$, $b_i^\dagger (g)$ is obtained by a Bogoliubov transformation from the $b_i, b_i^\dagger$ in the rotationally invariant case. Equivalently, the wavefunction can be expressed by a unitary transformation $\Psi_L(g)=\exp(-i\xi_{\alpha\beta}\Lambda^{\alpha\beta})\Psi_L(0)$, where $\xi_{\alpha\beta}$ is a real symmetric tensor and $\Lambda^{\alpha\beta}=\frac{1}{4}\sum_i \{ R_i^a, R_i^b\}$ is the generator of area-preserving diffeomorphisms~\cite{duncan}. The expression for the transformation matrix and the first-quantized expression for the wavefunction in Eq.(\ref{laughlinwfg}) is given in Ref.~\onlinecite{kunyang_aniso}.

The freedom in choosing $g$ implies that the usual rotationally-symmetric Laughlin wavefunction is a representative of a class of wavefunctions. 
However, being a topological phase, the physics of the Laughlin state does not 
depend on any given metric or lengthscales. Various wavefunctions $\Psi_L^{\nu=1/m} (g)$ differ from one another microscopically in terms of the shapes of their elementary droplets. For a FQH state at filling $\nu=p/q$ ($p$, $q$ are not necessarily co-prime), an elementary droplet is a unit of fluid containing $p$ particles in an area that encloses $q$ flux quanta.
The incompressible state is a condensate of such elementary droplets. For example, at $\nu=1/3$ we have a single particle 
occupying each three consequtive orbitals and preventing more particles from populating this region. In the language of root partitions and the Jack polynomials~\cite{jack}, the Laughlin $\nu=1/3$ state is defined by a root pattern $100100100100100\ldots$, and therefore its elementary droplet
is $100$. Note that these simple patterns only serve as labels for correlated wavefunctions that cannot be thought of as a simple crystal of electrons pinned at each third orbital and repelling each other via electrostatic forces. 

\begin{figure}[ttt]
 \includegraphics[width=0.85\linewidth]{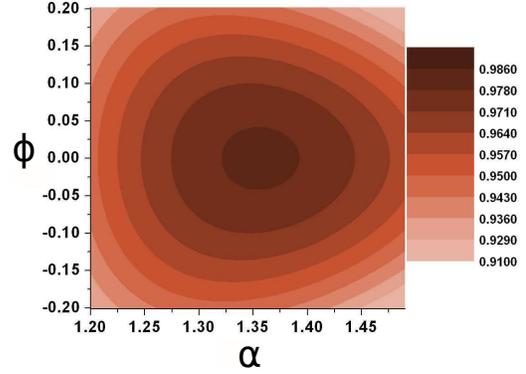}
\caption{(Color online). Overlap between the Coulomb ground state at $\nu=1/3$ for fixed anisotropy $\alpha_0=2,\phi_0=0$ and the family of Laughlin states parametrized by varying $\alpha$, $\phi$. The system is $N_e=9$ electrons on a hexagonal torus.}
\label{fig_optimize}
\end{figure}
For a model wavefunction, the ``gauge" freedom in $b,b^\dagger$ implies that the shape of elementary droplets changes with varying $g$, but the basic physical properties remain invariant unless the anisotropy magnitude $\alpha$ becomes too large or too small. If $\alpha$ is such that the maximum effective separation between electrons along some direction is of the order of $\ell_B$ or smaller, the FQH liquid correlations are expected to break down and CDW states might be favored (we will present some examples of this in Sec.~\ref{sec_ll1}). On the other hand, a generic state involves a competition of two different metrics -- the cyclotron metric, $g_m$, and the Coulomb metric $g_C$, therefore it is best approximated by the Laughlin state with an intrinsic metric $g$ that is generally different from both $g_m$ and $g_C$. Intrinsic metric $g$ is the one that minimizes the variational energy $E_g$,
\begin{equation}\label{correnergy}
E_g = \frac{\langle \Psi_L(g) | H(g_m, g_C)  | \Psi_L (g) \rangle}{\langle \Psi_L(g) | \Psi_L (g) \rangle}.
\end{equation}

To find the intrinsic metric in the microscopic calculation, we use a slightly different criterion that $g$ should maximize the overlap with the exact ground state of $H(g_m,g_C)$. In the rotationally-invariant case, the ground state of the Coulomb interaction at $\nu=1/3$ is known to have a remarkably high overlap with the Laughlin wavefunction. The overlap is defined as a scalar product between two normalized vectors, and in this particular case it is typically greater than 97\%. Therefore, we expect the intrisic metric chosen to maximize the overlap also to minimize the correlation energy (\ref{correnergy}). 
To obtain the anisotropic Laughlin states, we perform exact diagonalization of the ``$V_1$ Hamiltonian" on the torus. This Hamiltonian gives $\Psi_L^{\nu=1/3}$ as a unique and densest zero-energy ground state~\cite{haldane_prange}. Note that any translationally-invariant interaction can be expanded in terms of the Laguerre polynomials, $V(\mathbf{q})=\sum_m V_m \mathcal{L}_m(\mathbf{q}^2)$, where the coefficients $V_m$ are the Haldane pseudopotentials~\cite{haldane_prange}. Truncating this expansion at the first term, $V_1 \mathcal{L}_1(\mathbf{q}^2)$, singles out the strongest (``hard-core") component, which defines the Laughlin state at filling $\nu=1/3$. As the pseudopotential Hamiltonian is just a projection operator in the relative angular momentum space, the metric in $V_1 \mathcal{L}_1(\mathbf{q}^2)$ is the same as that originating from the cyclotron orbits.

\begin{figure}[ttt]
 \includegraphics[width=0.8\linewidth,angle=0]{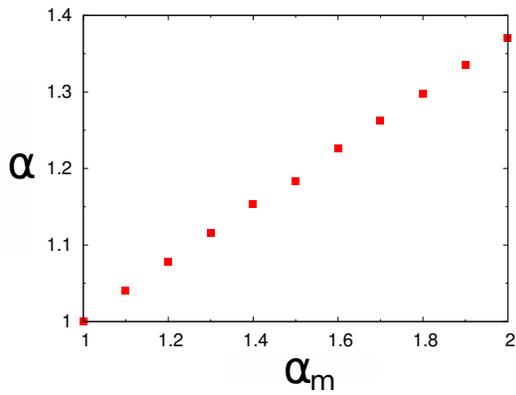}
\caption{Dependence of the intrinsic metric $\alpha$ on the mass metric $\alpha_m$ (Coulomb metric is set to identity).}
\label{fig_intrinsic}
\end{figure}
In Fig.~\ref{fig_optimize} we pick the ground state of the Coulomb interaction with fixed mass anisotropy $\alpha_0=2, \phi_0=0$ (the metric of the dielectric tensor is implicitly assumed to be $\alpha=1, \phi=0$), and we evaluate the overlap with a family of Laughlin states generated by varying $\alpha, \phi$. The overlap $|\langle \Psi_L^{\alpha,\phi}| \Psi_C^{\alpha_0=2,\phi_0=0}\rangle|$ is plotted as a function of $\alpha$ and $\phi$. We observe that the principal axis of the Laughlin state is aligned with that of the Coulomb state (maximum overlap occurs for $\phi=\phi_0=0$). Interestingly, the maximum overlap does \emph{not} occur for $\alpha=\alpha_0$, but for some value of the anisotropy that is a ``compromise" between the dielectric $\alpha=1$ and a cyclotron one $\alpha=2$. The value of the anisotropy that defines the intrinsic metric depends linearly on the band mass anisotropy (Fig.~\ref{fig_intrinsic}). This result illustrates the ability of the Laughlin state to optimize the shape of its fundamental droplets and maximize the overlap with a given anisotropic ground state of a finite system. 

\begin{figure}[ttt]
 \includegraphics[width=1\linewidth,angle=0]{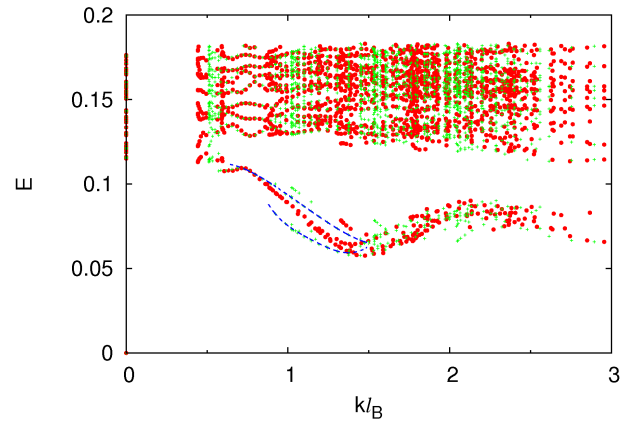}
\caption{(Color online). Energy spectrum of $N_e=9$ electrons at $\nu=1/3$ with the effective mass anisotropy $\alpha_m=2$  along the $x$-axis. When plotted as a function of $\sqrt{k_x^2+k_y^2}$ (green crosses), two branches of the magneto-roton mode are present (blue dotted lines are guide to the eye). If the spectrum is plotted as a function of $\sqrt{g^{ab}k_a k_b}$, the two branches collapse onto the same curve.}
\label{fig_rotonaniso}
\end{figure}

An alternative way to obtain the intrinsic metric is to analyze the shape of the lowest excitation -- the magneto-roton mode, which was successfully described by the single-mode approximation~\cite{sma}.  In a rotationally-invariant case, this mode has a minimum at $k \sim \ell_B^{-1}$. In the presence of anisotropy, the minimum occurs at different $|k|$ in the different directions (Fig.~\ref{fig_rotonaniso}). This leads to an alternative definition of the intrinsic metric based on the shape of the roton minimum in the 2D momentum plane. We numerically establish that this definition agrees well with our previous definition of the intrinsic metric.
In Fig.~\ref{fig_rotonaniso} we plot the energy spectrum of an anisotropic Coulomb interaction at $\nu=1/3$ as a function of the rescaled momentum $\sqrt{g^{ab}k_a k_b}$, where $g$ is the guiding center metric that maximizes the overlap with the family of Laughlin wavefunctions (Fig.~\ref{fig_intrinsic}). With the usual definition of the momentum $|k|$, several roton minima appear. Different magneto-roton branches collapse onto the same curve if we plot them as a function of $\sqrt{g^{ab}k_ak_b}$. This is reasonable, because the magneto-roton mode is well approximated by single-mode approximation up to the roton minimum~\cite{boyang}, which is defined entirely in terms of the properties of the ground state. The anisotropy of the ground state structure factor (determined by the shape of elementary droplets) dictates the position of the roton minimum.

The analysis of this section in principle applies to other LLL states at fillings $\nu=p/(2p+1), p=2,3,\ldots$, though it is more involved because of the ``multicomponent nature" of these states and typically a smaller excitation gap. 

\section{Higher Landau levels: quantum phase transitions driven by anisotropy} \label{sec_ll1}

We have found that $\nu=1/3$ in the LLL is particularly robust with respect to anisotropy, and this is the case also with other prominent FQH states. In higher LLs, due to a number of nodes in the single-particle wavefunction, the region of the phase diagram where incompressible states occur becomes increasingly narrower, and compressible phases such as stripes and bubbles take over. In this Section we discuss the effects of anisotropy on FQH states in higher LLs, focusing on fillings $\nu=1/3$ and $1/2$. Because of closer energy scales, we find that moderate changes in the anisotropy induce phase transitions between compressible and incompressible phases. 

\subsection{$n \geq 2$ Landau levels: stripes and bubbles}\label{ll2}

\begin{figure}[ttt]
 \includegraphics[width=0.95\linewidth,angle=0]{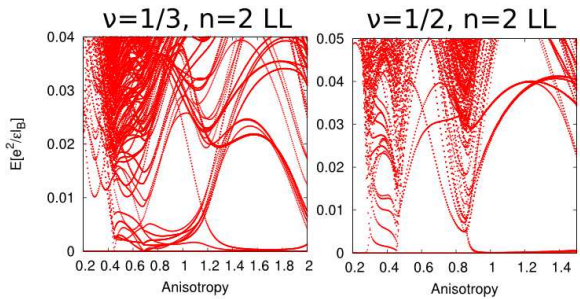}
\caption{(Color online). Energy spectrum of $\nu=1/3$ (left) and $\nu=1/2$ filled $n=2$ LL (right): mass anisotropy establishes and reinforces the stripe
order.}
\label{fig_ll2}
\end{figure}  
In $n=2$ LL and higher, isotropic FQH states are energetically less favorable than stripe and bubble phases at filling $\nu=n+1/2$ and $\nu=n+1/3$, respectively. In Fig.~\ref{fig_ll2} we show the energy spectrum (in units of $e^2/\epsilon\ell_B$) as a function of anisotropy $\alpha$ (we set the angle $\phi$ to zero). Energies are plotted relative to the ground state at each $\alpha$, and we chose the relatively modest system sizes ($N_e=8$ and 10 electrons) to facilitate comparison with the existing isotropic data in the literature~\cite{edduncanky_stripe, edduncanky_bubble}. The aspect ratio is set to the optimal values for the appearance of stripes or bubbles (see Refs.~\onlinecite{edduncanky_stripe} and~\onlinecite{edduncanky_bubble}).

As we see on the right panel of Fig.~\ref{fig_ll2}, at $\nu=1/2$ the presence of mass anisotropy reinforces the stripe when $\alpha$ is increased. This leads to a more pronounced quasi-degeneracy of the ground-state multiplet, and an increase of the gap between this multiplet and the excited states. For yet larger values of $\alpha$, it appears that some of these excited states may become the ground state, however this occurs for very large $\alpha$ when this finite system effectively becomes one-dimensional under anisotropy deformations. 

In case of $\nu=2+1/3$ state, it has been argued that the isotropy point is described by a two-dimensional CDW order known as the bubble phase~\cite{edduncanky_bubble}. A bubble differs from a stripe in having a larger degeneracy and a two-dimensional mesh of (quasi)degenerate ground-state wavevectors (as opposed to the one-dimensional array in case of a stripe). The spread of the quasidegenerate levels was also found to be somewhat larger than in case of stripes. All of these features are obvious in Fig.\ref{fig_ll2} (left) for $\alpha=1$. The bubble phase remains stable to some extent when $\alpha$ is reduced; for very small $\alpha$ it is eventually destroyed and replaced by a simple CDW. On the other hand, when $\alpha$ is increased, a smaller subset of momenta becomes very closely degenerate with some of the excited levels. This second-order (or weakly first order) transition results in a stripe phase. As for the $\nu=1/2$ case, this stripe becomes enhanced as $\alpha$ is further increased.  Therefore, in $n\geq 2$ LLs mass anisotropy generally produces stripes, even when isotropic ground states have a tendency to forming a bubble phase.   

\begin{figure}[ttt]
 \includegraphics[width=0.9\linewidth,angle=0]{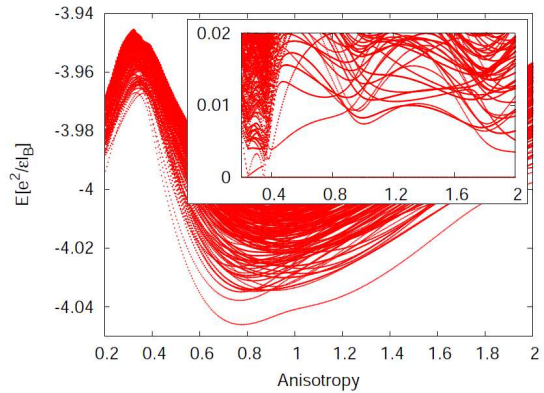}
\caption{(Color online). Spectrum of $N_e=8$ electrons at $\nu=1+1/3$ with thickness $w=2\ell_B$. Inset: same spectrum plotted relative to the ground state at each $\alpha$. Unit cell has a rectangular shape with aspect ratio $3/4$.}
\label{fig_onethirdanisoll1w2.0}
\end{figure}

\subsection{$n=1$ Landau level: incompressible to compressible transitions driven by anisotropy}

In $n=1$ LL, $\nu=1/3$ state is significantly weaker than its $n=0$ LL counterpart, having an experimental gap an order of magnitude smaller and roughly the same as the gap of $\nu=1/2$
state. This has been anticipated in early numerical calculations that found the ground state of the Coulomb interaction projected to $n=1$ LL to be at the transition point between compressible and incompressible phases~\cite{haldane_prange}.  

\begin{figure}[ttt]
 \includegraphics[width=0.8\linewidth]{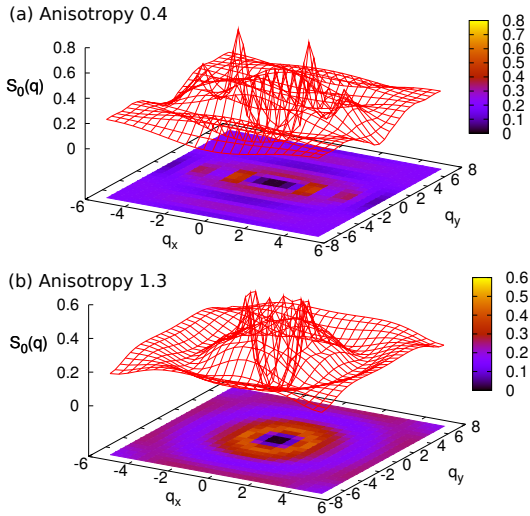}
\caption{(Color online). Guiding-center structure factor $S_0(\mathbf{q})$ for $\nu=1/3$ state in $n=1$ LL with thickness $w=2l_B$ and anisotropy $\alpha=0.4$ (a). For comparison, we also show $S_0(\mathbf{q})$ for the state with $\alpha=1.3$ which is in the Laughlin universality class (b). Two peaks in the response function (a) represent the onset of compressibility and CDW ordering.}
\label{fig_s0qonethird}
\end{figure}  

Although idealized numerical calculations with pure (projected) Coulomb interaction work exceedingly well in $n=0$ LL, more realistic models are required to describe phases in $n=1$ LL. In particular, the inclusion of finite width effects~\cite{peterson} and varying a few strongest Haldane pseudopotentials in necessary to determine the phase diagram. We find that varying the $V_1$ pseudopotential leads to the following outcomes: (i) generically, for $\delta V_1<0$, the system is pushed deeper into a compressible phase; (ii) for $\delta V_1>0$, finite-size calculations on systems up to $N_e=9$ electrons permit the existence of two regimes: for $0<\delta V_1^a < \delta V_1 < \delta V_1^b$, the ground state is in the Laughlin universality class, but the lowest excitation is not the magneto-roton; for $\delta V_1 > \delta V_1^b$, the ground state \emph{and} the excitation spectrum is the same as in $n=0$ LL. For smaller systems, $\delta V_1^b$ is estimated to be around $0.1e^2/\epsilon \ell_B$, while $\delta V_1^a$ is around $0.04e^2/\epsilon \ell_B$. Larger systems suggest that these two points might merge in the thermodynamic limit, when only a small modification of the interaction might be needed for the Laughlin physics to appear at $\nu=1/3$ in $n=1$ LL. Alternatively, we can consider the Fang-Howard ansatz that mimicks the finite-width effects. In this case, the width of $\ell_B$ or smaller is sufficient to drive a phase transition between the compressible state and the Laughlin-like state, in agreement with results on the sphere and using an alternative finite-width ansatz~\cite{papic_zds}.  

In summary, the ground state at $\nu=1+1/3$ very likely belongs to the Laughlin universality class. We note that the collective mode in this case displays significantly more wiggles than in the LLL (some wiggles exist in case of $n=0$ LL Coulomb state, but they are less pronounced). For large momenta, the magneto-roton mode also appears to merge with the continuum of quasiparticle-quasihole excitations. This is likely a finite-size artefact, although we cannot rule out that it represents an intrinsic feature of $\nu=1+1/3$ state, in which case it might have an observable signature in optical experiments that distinguishes it from $\nu=1/3$ state.  

Because of the fragility of $\nu=1+1/3$ state, we expect that mass anisotropy might have more dramatic consequences than in the LLL. In Fig.~\ref{fig_onethirdanisoll1w2.0} we plot the energy spectrum as a function of anisotropy. One notices that the isotropy point ($\alpha=1$) does not bear any special importance -- indeed, the system appears more stable in the vicinity of it where it can lower its ground state energy or increase the neutral gap. On either side of the isotropy point, however, the system remains in the Laughlin universality class; e.g. at $\alpha=0.8$ and $\alpha=1.3$ the maximum overlap with the Laughlin state is $75\%$ and $80\%$, respectively (these overlaps, although modest compared to the standards of $n=0$ LL, can be adiabatically further increased by tuning the $V_1$ pseudopotential). Note that the quoted maximum overlaps are achieved by the Laughlin state with $\alpha'$ somewhat different from $\alpha$ of the Coulomb state, analogous to Fig.\ref{fig_optimize}. 

The new aspect of Fig.\ref{fig_onethirdanisoll1w2.0} is the transition to a compressible state with CDW ordering for $\alpha \lesssim 0.4$. In that region of parameter space, the system is very sensitive to changes in the boundary condition -- the sharp degeneracies seen in rectangular geometry in Fig.\ref{fig_onethirdanisoll1w2.0} are not obvious in case of higher symmetry, square or hexagonal, unit cell. As an additional diagnostic tool for the compressible states, it is useful to consider a guiding-center structure factor,
\begin{equation}\label{sq}
S_0(\mathbf{q}) = \frac{1}{N_\phi} \sum_{i,j}\langle e^{i\mathbf{q}\cdot \mathbf{R}_i} e^{-i\mathbf{q}\cdot \mathbf{R}_j} \rangle - \langle e^{i\mathbf{q}\cdot \mathbf{R}_i} \rangle \langle e^{-i\mathbf{q}\cdot \mathbf{R}_j} \rangle,
\end{equation}
where the expression for the Fourier components of the guiding-center density, $\rho (\mathbf{q}) = \sum_i^N e^{i\mathbf{q}\cdot \mathbf{R}_i} $,
has been used. Note that $S_0(\mathbf{q})$ is normalized per flux quantum rather than (conventional) per particle~\cite{duncan}. In Fig.\ref{fig_s0qonethird}(a) we show the plot of $S_0(\mathbf{q})$ evaluated for the state with $\alpha=0.4$ in Fig.\ref{fig_onethirdanisoll1w2.0}. Two sharp peaks in the response, similar to those previously identified in $n\geq 2$ LL states~\cite{edduncanky_stripe}, are the hallmark of CDW order. They are to be contrasted with the smooth response in case of an anisotropic state in the Laughlin universality class for $\alpha=1.3$, Fig.\ref{fig_s0qonethird}(b).

\begin{figure}[ttt]
 \includegraphics[width=0.78\linewidth,angle=270]{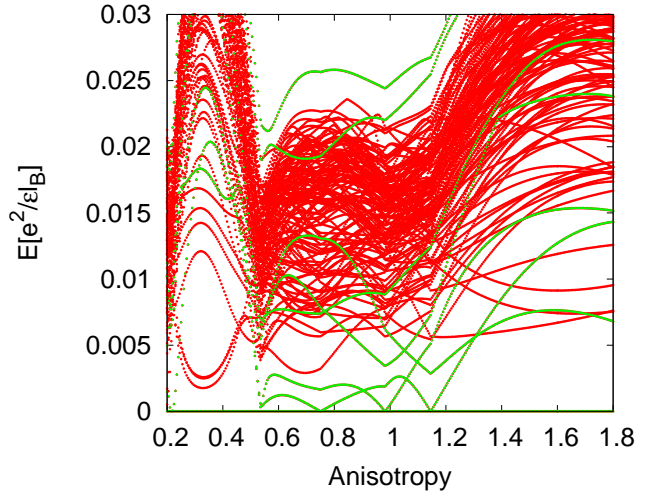}
\caption{(Color online). Spectrum of $N_e=14$ electrons at $\nu=1+1/2$ with thickness $w=2l_B$, as a function of anisotropy $\alpha$. Energies are plotted relative to the ground state at each $\alpha$, and the unit cell has a rectangular shape with aspect ratio $3/4$.}
\label{fig_10_20}
\end{figure}

As a second example in $n=1$ LL, we consider half filling where the Moore-Read Pfaffian state~\cite{mr} is believed to be realized in some regions of the phase diagram. This state has a non-Abelian nature, which is reflected in the non-trivial ground state degeneracy~\cite{readgreen} when subjected to periodic boundary conditions. For $\nu=1/2$, the eigenstates of any translationally-invariant interaction possess a twofold center-of-mass degeneracy~\cite{duncan_translations}. On top of this, Moore-Read state has an additional threefold degeneracy. Conventionally, the many-body Brillouin zone is defined for $p=1$,$q=2$ and has a size $N^2$ ($N$ being the GCD of $N_e$ and $N_\phi$), which forces the degenerate groundstates to belong to a Brillouin zone corner $\mathbf{K}=(N/2,N/2)$ and centers of the sides, $\mathbf{K}=(0,N/2);(N/2,0)$. It is also possible to define a ``quartered" Brillouin zone such that the three degenerate states are all mapped to zero momentum~\cite{zed}. 
The three $\mathbf{K}$ sectors are equivalent for a hexagonal unit cell, however in an anisotropic system the degeneracy is always lifted.

In Fig.\ref{fig_10_20} we plot the spectrum of the Coulomb interaction as a function of anisotropy (states belonging to $\mathbf{K}$ sectors where the Moore-Read state is realized, are indicated). As earlier, we assume finite width of $w=2\ell_B$ in order to instate the Pfaffian correlations~\cite{rh_pf}. Note that our calculation only uses two-body (Coulomb) interaction, therefore in each finite system the Moore-Read state will mix with its particle-hole conjugate pair, the anti-Pfaffian~\cite{antipf}. The mixing between the two states can be controlled by including higher LLs~\cite{llmix}.  For $0.5 \leq \alpha \leq 1.3$, we find a three-fold quasi-degenerate multiplet, suggesting the presence of Moore-Read state at the isotropy point and in the neighborhood of it. In finite systems, there is some splitting of the degeneracy that might be reduced upon tuning the $V_1,V_3$ pseudopotentials. Also, upon tuning the anisotropy around $\alpha=1$, there are crossings within the multiplet of degenerate ground states without apparent closing of the gap. 
The region of the Moore-Read state is defined by sharp transitions towards crystal phases. These transitions are likely second order because they do not appear to involve any level crossing, but rather lifting of the degeneracy within a ground-state multiplet.

\section{Conclusion} \label{sec_conclusion}

We have presented a method to study the effects of anisotropy on FQH phases in finite-size systems. We found that the prominent FQH states (as in the lowest Landau level) are robust to variations of anisotropy, due to the adjustment of the intrinsic metric describing the shape of their elementary droplets. As we demonstrated using the example of the Laughlin $\nu=1/3$ state, this metric is usually a compromise between the metric dictated by the cyclotron motion and the metric originating in the dielectic environment of the 2DES. In this sense, it is unlike the non-interacting Landau level problem, or the problem of weak localization~\cite{wolfle}, where the anisotropy can be completely ``gauged away" (i.e. removed) by length rescaling. Instead, it is more akin to the problem of shallow donors in many-valley semiconductors~\cite{kohn}. Indeed, such compromise picture leads to a quantitatively accurate description of the variation of the critical density of the metal-insulator transition (an intrinsically many-body phenomenon) in three-dimensional doped many-balley semiconductors~\cite{ravin}, so one may wish to ascertain to what extent this can lead to quantitative predictions in the FQHE case. In higher LLs, anisotropy induces quantum phase transitions, likely of second order, to compressible phases with broken symmetry.

Anisotropy is an important aspect of FQHE as it represents a mechanism that probes the intrinsic metric of incompressible fluids in the geometrical picture of the FQHE. In addition, because our calculations show the possibility of phase transitions in the $n>0$ Landau levels as a function of mass anisotropy, it motivates experimental studies on systems with both moderate mass anisotropy (e.g. AlAs and Si, $\alpha \sim 3-5$), as well as systems with large mass anisotropy (e.g. Ge, $\alpha\sim 20$), where behavior may be different in the upper Landau levels from the anisotropic GaAs. In these systems, as in GaAs, anisotropy could be furher tuned using tilted fields, thereby adding to the richness of the FQHE phenomena. 

\section{Acknowledgments}

This work was supported by DOE grant DE-SC$0002140$.

\end{document}